\documentstyle[12pt]{article}
\begin{document}
\vskip 2 cm
\begin{center}
\Large{\bf SIGNALS OF SUPERSYMMETRIC DARK MATTER } 
\end{center}
\vskip 2 cm 
\begin{center}
{\bf AFSAR ABBAS}
\vskip 5 mm
Institute of Physics, Bhubaneswar-751005, India

(e-mail : afsar@iopb.res.in)
\end{center}
\vskip 20 mm
\begin{centerline}
{\bf Abstract }
\end{centerline}
\vskip 3 mm
The Lightest Supersymmetric Particle predicted in most of
the supersymmetric scenarios is an ideal candidate for the
dark matter of cosmology. Their detection is of extreme significance
today. Recently there have been intriguing signals of a 59 Gev neutralino
dark matter at DAMA in Gran Sasso. We look at other possible signatures
of dark matter in astrophysical and geological frameworks.
The passage of the earth through dense clumps of dark matter
would produce large quantities of heat in the interior of 
this planet through the capture and subsequent annihilation of dark
matter particles. This heat would lead to large-scale volcanism
which could in turn have caused mass extinctions. The periodicity 
of such volcanic outbursts agrees with the frequency of 
palaeontological mass extinctions as well as the observed 
periodicity in the occurrence of the largest flood basalt 
provinces on the globe. Binary character of these extinctions
is another unique aspect of this signature of dark matter.
In addition dark matter annihilations 
appear to be a  new source of heat in the planetary systems.
\newpage
Careful measurements on the dynamics of galaxies by Fritz Zwicky back 
in the 1930's led him to the conclusion that the mass to light ratio of
galaxies and clusters of galaxies required far more mass than explained on
the basis of stellar origin of their light.
Hence he argued for the existence of invisible dark matter. His ideas were
not immediately appreciated and it took several decades for astronomers
and physicists to understand the significance of this discovery.
Today, on the basis of several experimental observations, it has become
clear that to account for the observed motion in the cosmos, gravitational
fields much stronger than those attributable to luminous matter are
required. As much as $ 90 \% $ of the mass in the universe is
constituted of this invisible dark matter.

This conclusion gets further support from simulations using cosmological
models which bring out the necessity for large number of relic particles
from the early universe. The ideal candidate for these relic species are
the weakly interacting massive particles ( WIMPS ). These WIMPS arise
most naturally in Supersymmetric theories.

Most of the supersymmetric theories contain one stable particle , the
so called Lightest Supersymmetric Particle ( LSP ), which is the candidate
for dark matter as a WIMP. The existence of a stable supersymmetric
partner particle results from the fact that these models include a
conserved multiplicative quantum number, the R-parity.
This takes on values of +1 and -1 for particle and supersymmetric partners
respectively. As per this conservation principle SUSY particles can only
be generated in pairs.This requires that SUSY particles may decay in odd
number of particles only. As such the LSP must be stable.

However the R-parity may be violated.
The quantum number R is given by
\begin{equation}
R  =  ( -1 ) ^{3 B + L +2 S}
\end{equation}
where B is baryon number, L is lepton number and S is the spin. A
violation of B or L implies a violation of R number. However sharp
bounds on the violation of R have been established.

Had LSP been susceptible to the strong or the electromagnetic
interactions, then it would have been detectable today, as it would have 
condensed with ordinary matter. Bounds on the abundance of LSP
normalized with respect to the abundance of proton have been calculated
[1]
\begin{equation}
{n (LSP) \over {n(p)}} \sim 10^{-10} (strong) \ldots 10^{-6}
(electomagnetic)
\end{equation}

Hence LSP should be practically bereft of the
strong or the electromagnetic interactions. It can however take part
in gravitational and weak interactions. So these are ideal candidates for
the WIMP scenarios of the cosmological dark matter.
SUSY dark matter particles are of interest as they occur in a totally
different context and not specifically introduced to solve the
dark matter problem.

Possible candidates for the LSP include the photino ( S=1/2 ),
the higgsino ( S=1/2 ), the zino ( S=1/2 ), the sneutrino ( S=0 )
and the gravitino ( S=3/2 ).
The above spin 1/2 SUSY particles are called gauginos. In most of the 
recent theories the favourite LSP is the neutralino which is
defined as the lowest mass linear
superposition of photino ($\tilde\gamma$), zino ($\tilde {Z}$)
and the two higgsino states ( $\tilde {H_1}$ , $\tilde {H_2}$ ) :

\begin{equation}
\chi = a_1 {\tilde\gamma} + a_2 {\tilde {Z}}
       + a_3 {\tilde {H_1}} + a_4 {\tilde {H_2}}
\end{equation}

Within the Minimal Supersymmetric extension of the Standard Model
( MSSM ) , it is convenient to describe the supersymmetry phenomenology at
the electroweak scale without too strong theoretical assumptions.
Various properties like relic abundances and detection rates have been
carefully analyzed recently by several authors [2]. In the
soft-breaking Lagrangian one has the trilinear and bilinear breaking
parameters.

Either looking for signatures of the dark matter or detecting it directly
is obviously, a major enterprise today [1]. To understand the properties
of the
elusive and invisible dark matter, one may try to detect the dark matter
directly or look for situations where it would have left
its indelible fingerprints. The latter would be referred to as indirect
detection.

First the direct detection. The most exciting news is that recently
there have been intriguing tell-tale signs of dark matter. There are
several detectors all over the world trying to catch a dark matter
particle, Most of them focus on WIMP-nucleus elastic scattering from
target nuclei part of the detector. The putative WIMP would be
detected via nuclear recoil energies which are expected to be in the
kilo-electronvolt range. 

The experiment, which has found possible signature of dark matter, is
DAMA, which is housed deep underground in the INFN Gran Sasso
National Laboratory in Italy [3]. It this detector high atomic-number 
target nuclei, such as Iodine ( in the form of NaI) and Xenon are used.
To help isolate a possible WIMP signal from the background, one
focuses on the annual modulation effect. As the earth rotates
around the sun, the dynamics are such that its rotational velocity 
would be in the same direction as that of the solar system with
respect to the galaxy in June and opposite in December.
This would bring in an annual modulation in the WIMP detection
rate. The 100 kg DAMA detector, after two years of
data collection on this modulation effect, has enabled the experimental
group to announce the possible detection of a 59 Gev WIMP, most
likely a neutralino [3].
This is a most significant discovery in the direct dark matter
detection set-ups. Further work continues to be done to
consolidate or refute this discovery.

One may ask for possible  indirect
signatures of the dark matter in the universe. One has to seek
out special and unique scenarios in the astrophysical or geological
context to obtain unique signatures of dark matter. A few such scenarios
studied by us are described below [4,5].

While investigating the possibility that a WIMP could explain
both the dark matter problem and the solar neutrino problem,
Press and Spergel [6]    
estimated  the rate at which the sun or a planet will capture WIMPs.
As given by Krauss et al the capture rate for earth is [7] :

\begin{equation}
 \dot{N_{E}} = ( 4.7 \times 10^{17} sec^{-1} ) 
         \{ 3ab \}  
         \left[ \frac { \rho_{0.3} \sigma_{N,32} }
                      { \bar{v}_{300}^{3} } 
                \right] 
         ( \frac {1}{ 1 + m_{X}^{2} / m_{N}^{2} } ) \\
\end{equation}
 
\noindent where 
$ m_X $ is the mass of the DM particle, 
$ m_N $ is the mass of a typical nucleus off which the the particle
  elastically scatters with cross-section $ \sigma_N $,
$ \rho_X $ is the mean mass density of DM particles in the 
  Solar System,
$ \bar{v} $ is the r.m.s. velocity of dark matter in the Solar System, 
$ \rho_{0.3} =  \rho_X / 0.3 GeV cm^{-3}  $,
$ \sigma_{N,32} =  \sigma_N / 10^{-32} cm^2  $,
$ \bar{v}_{300} =  \bar{v} / 300 km s^{-1}  $, 
and
$ a $ and $ b $ are numerical factors of order unity which depend on the
density profile of the sun or planet. 

The earth will continue to accrete more and more particles until their
number density inside the planet becomes so high that they start to
annihilate. One possibility is a flux of upwardly moving neutrinos at the 
earth's surface. This has been studied very meticulously and is being
used to detect dark matter directly [7-9].
We ignore this channel and study other possible outcomes of the said
annihilation of dark matter at the centre of earth.
This had not been studied earlier.

Depending on whether the dark matter is neutralino, photino, 
gravitino, sneutrino, Majorana neutrino or some other, different 
annihilation channels are possible [8-10]. Note that we are however,
looking in particular, at neutralino in the supersymmetry broken
scenario of the MSSM as described above [2]. 
Generally the most significant channels are
$ \chi \bar{\chi} \rightarrow  q \bar{q} $ ( quark-antiquark ), 
$ \chi \bar{\chi} \rightarrow \gamma \gamma $ ( photons ) and
$ \chi \bar{\chi} \rightarrow l \bar{l} $ ( lepton-antilepton ).  

We ignore the $\nu \bar{\nu}$ which has been well studied by others
[8-10] and concentrate upon photon producing channels.
In the quark channel hadronization will take place through jets and
subsequent radiative decay will lead to mesons which in turn will decay 
through their available channels. Hence
[10] :
\begin{equation}
\chi \bar{\chi} \rightarrow q \bar{q} \rightarrow 
  ( \pi^{0}, \eta, ... ) 
  \rightarrow \gamma + Y  
\end{equation} 

\noindent All annihilation processes 
which directly or indirectly create
photons, energy is delivered to the core through inelastic 
collisions. This would lead to the generation of heat in the earth's core.
We wish to study this heat generation in the core
through annihilation. This heat is :
\begin{equation} 
 \dot{Q}_{E} = e \dot{ N_E } m_X 
\end{equation} 

\noindent where 
$ e $ is the fraction of annihilations which lead to the generation of 
heat in the core of the earth.
Here e may be as large as unity for the ideal case where
the WIMPs annihilate predominantly through photons only.
For an order of magnitude estimate
let us take it to be $ \sim 0.5 $ [8-10]. 

On taking $ ab \sim 0.34 $ 
[2],
$ \rho_{0.3} =1 $, 
$ \bar{v}_{300} = 1 $, 
$ m_{X} = 55 GeV $
and the cross-section on iron to be $ \sigma_N = 10^{-33} cm^2 $,
one finds that $ \sim 10^8 W $
of heat is generated. 

As the visible matter clumps together to form stars, planets, etc. an
interesting question is whether the dark matter also displays this
tendency of clumping. Interestingly several dark matter models do
suggest that clumps of dark matter arise naturally during the course of
evolution of the universe.  Silk and Stebbins [11] considered cold
dark matter models with cosmic strings and textures appropriate for galaxy
formation.  They found that a fraction $ 10^{-3} $ of the galactic halo
dark matter may exist in the form of dense cores. These may survive up to
mass scales of $ 10^8 M_{\odot} $ in galaxy halos and globular clusters
[11]. Analyzing the stability of these clumps of dark matter, they
found that the cores of these clumps will not be affected, although the
outer layers may be stripped off by tidal forces. In the cosmic string
model, the clumpiness C, defined as the ratio of clumped matter
concentration to normal concentration, of dark matter at the present epoch
would be [11]

\begin{equation}
C \sim 10^{12} f_{cl} h^6 \Omega_{0}^3
\end{equation}

where $ f_{cl} $ is the fraction of dark matter in clumps, 
$ H $ is the Hubble parameter parametrized as 
$ 100 ~h ~km/s ~Mpc^{-1} $, and 
$ \Omega_{0} $ is the closure energy density of the Universe.  

Subsequently
Kolb and Thachev [12] studied isothermal fluctuations in the dark
matter density during the early universe. If the density of the isothermal
dark matter fluctuation or clumps, 
$ \Phi = \delta \rho_{DM} / \rho_{DM} $,
exceeds unity, a fluctuation collapses in the radiation-dominated epoch
and produces a dense dark matter object. They found the final density
of the virialized object $ \rho_{F} $ to be 
\begin{equation}
\rho_{F} \sim 140 \Phi^{3} ( \Phi + 1 ) \rho_{x}
\end{equation}
where $ \rho_{x} $ is equilibrium density.

For axions, a putative dark matter particle,
density fluctuations can be very high, possibly spanning the
range $ 1 < \Phi < 10^4 $. The resultant density in miniclusters can be as
much as $ 10^{10} $ times larger than the local galactic halo density. The
probability at present of an encounter of the earth with such an axion
minicluster is 1 per $ 10^7 $ years with $ \Phi = 1 $. Kolb and Tkachev
found two types of axion clumps arising from two kinds of initial
perturbations:

\begin{itemize}
 \item{ Fluctuations with $ 10^{-3} < \Phi < 1 $ collapse in the
matter-dominated epoch. }
 \item{ Fluctuations with $ \Phi > 1 $ collapse in the radiation-dominated
epoch. }
\end{itemize}

  If the dark halo is mostly made of neutralinos, then the clumping factor
in the MSSM could be less than $ 10^9 $ for all neutralino masses [13].

It has been estimated that
these clumps would cross earth with a periodicity of 30-100
Myrs [14]. 
Thus during the passage of the earth through such clumps at 
regular intervals, the flux of the incident DM particles will
increase by roughly a factor of $\sim 10^9$. Consequently the value
of $ \dot{Q}_{E} $ during the passage 
of a clump will be $ \sim 10^{17} $ W [4].

Improving upon the previous work, Gould 
[15] 
obtained greatly enhanced capture rates for the earth ( 10-300 times
that previously believed ) when the WIMP mass roughly equals the
nuclear mass of an element present in the earth in large 
quantities, thereby constituting a resonant enhancement.
Gould's formula gives the capture rate for each element in the
earth as 
[15] : 

\begin{equation}
\dot{ N_{E} } = ( 4.0 \times 10^{16} sec^{-1} )
         \bar\rho_{0.4}   
         \frac { \mu } { \mu_{+}^{2} }
         Q^{2}
         f
         \left< \hat\phi 
                ( 1- \frac { 1-e^{-A^2} } { A^2 } )
                \xi_{1} (A)
                \right>
\end{equation}

\noindent where
$ \bar\rho_{0.4} $ is the halo WIMP density normalized to 
   $ 0.4 GeVcm^{-3} $ ,
$ Q = N - ( 1 - 4 sin^2 \theta_W  ) Z $
   $\sim$ N - 0.124Z,
$ f $ is the fraction of the earth's mass due to this element,   
$ A^2 = ( 3 v^2 \mu ) / ( 2 \hat{v}^2 \mu_{-} ) $,
$ \mu =  m_{X} / m_{N}  $,
$ \mu_{+} = ( \mu + 1 ) / 2  $,
$ \mu_{-} = ( \mu - 1 ) / 2  $,
$ \xi_{1} (A) $ is a correction factor,
$ v = $ escape velocity at the shell of earth material , 
$ \hat{v} = 3kT_w/m_X 
          = 300 kms^{-1} $ is the velocity dispersion, and 
$ \hat\phi =  v^2 / { v_{esc} }^2  $        
  is the dimensionless gravitational potential. 

  In the WIMP mass range 15 GeV-100 GeV this yields total
capture rates of the order of $ 10^{17} sec^{-1} $ to 
$ 10^{18} sec^{-1} $ .
According to the equation above, this yields
$\dot{Q}_{E} \sim 10^8 W - 10^{10} W $ for a uniform density 
distribution. 

  In the case of clumped DM with core densities $ 10^9 $
times the galactic halo density, global power production due
to the passage of the earth through a DM clump is
$ \sim 10^{17} W - 10^{19} W $. It is to be noted that this
heat generated in the core of the earth is huge and arises
due to the highly clumped CDM [4].

If the dark matter is composed of neutralinos, the effect of geological
heating may not be in the saturation regime [16] 
and this may diminish the heat production. The effect 
also depends on the unknown density inside the dark matter clump. The
estimates 
show that in case of the neutralino, it can reach the right order of 
magnitude for extreme values of parameters. One should note however, that 
not only are the parameters of neutralino interactions not well known,
but even the nature of the dark matter particles (neutralino or something
else?)
is not yet established. However, the bottom line is that our estimates
should be relevant for non-standard neutralino parameters and/or
other dark matter particles.

The geothermodynamic theory states that continuous 
heat absorption
by the the lowermost layer of the mantle, the so called D" layer would
result from a temporary increase in heat transfer from the core
[17].
This process would continue until, due to its 
decreasing density, this layer becomes
unstable, eventually breaking up into rising plumes.
This is the only physical possibility as plume production
is the most efficient way of heat transfer in earth.  
The lower mantle origin for plumes concept is 
strengthened by several recent observations. 
Firstly, high levels of primordial He-3 reported for Siberian flood 
basalts [18] 
support this view. Secondly, high levels of Osmium-187 
from the decay of the Rhenium-187, found in high concentrations in the
earth's core, observed in Siberian flood basalts [19], 
suggest that some of these rocks may even come from the
outer core.  

Due to its lower density,
a typical plume created in this manner would
well upwards. In this process, decompression of
the plume on account of its ascent in a pressure gradient will
lead to partial melting of the plume head, thereby producing 
copious amounts of basaltic magma [20].
Mantle velocities being 
$ \sim 1 $ m/year , such a plume would take $ \sim 5 $ million
years to reach the crust. It would then melt its way through 
the continental crust, thereby producing viscous acidic
( silicic ) magma [21].

The ultimate arrival of such a plume head at the surface could
be cataclysmic. Initial explosive silicic volcanism would be 
followed by periods of large-scale basalt volcanism that
ultimately lead to the formation of massive flood basalt
provinces such as the Siberian Traps, the Deccan Traps in India
and the Brazilian Paran\'{a} basalts.
Extensive atmospheric pollution would follow; the Deccan Trap
flood basalt volcanic episode ( $ \sim $ 65 million years ago )
ejected huge amounts of basalt, tonnes of $ H_2SO_4 $,  
$ HCl $, and fine dust [21]. 
Climatic models predict that this is capable of triggering
a chain of events ultimately leading to the depletion of the
ozone layer, global temperature changes, acid rain
and a decrease in surface ocean alkalinity.

Thus, Deccan volcanism has been proposed as a possible 
cause for the K/T ( Cretaceous/Tertiary ) mass extinction
that extinguished the dinosaurs   
[20,22], while
the Siberian basalts have been put forth as a 
possible culprit for the P/T ( Permian/Triassic )
mass extinction
[20]. In fact, there exists a striking concordance 
between the ages
of several major flood basalt provinces and the dates of
the major palaeontological mass extinctions
[17]. Hence it has been proposed by us [4] that all major
periodic mass extinctions have been caused by gigantic
volcanism which in turn were caused by the heat coming from 
dark matter annihilations at the centre of earth. So the actual culprit,
for all major extinctions including that of dinosaurs, was the
invisible dark matter [4,5].

Collar set forth the hypothesis that the 
passage of the clump leads to direct extinctions by causing cancers
in organisms [14]. If this is correct, then this extinction should 
precede that due to volcanism by approximately 5 million years. 
Hence each major extinction should, at higher 
resolution, be a binary extinction: the first extinction due to the direct
passage of the clump (causing cancers in various organisms),
ie. the carcinogenic dark matter scenario, and the
second extinction due to massive volcanism, ie. the volcanogenic dark matter 
scenario  above. What is the empirical situation regarding this
unique prediction of the dark matter extinction scenario ?
 
The Permo-Triassic extinction is the most severe ever recorded in the
history of life on earth. It has been estimated that 88 - 96 \% of all
species disappeared in the final stages of the Permian.
However, Stanley and Yang [23] discovered that this
biotic crisis in fact consisted of two distinct extinction events. The
first and less severe of the two was the Guadalupian crisis at the end of
the penultimate stage of the Permian, followed after an interval of
approximately 5 million years by the mammoth end-Tartarian event at the
P/T boundary.  Traditionally, the Signor-Lipps effect has been used to
explain the high rates of extinction during the last two stages of the
Permo-Triassic extinction. It was generally believed that the actual
extinction occurred at the Permo-Triassic boundary during the end of the
Tartarian stage, with the high Guadalupian metrics being due to the
`backward smearing' of the single grand extinction event. However, Stanley
and Yang found that the high rates of extinction of the Guadalupian stage
were not artifacts of the Signor-Lipps effect, but represent actual
extinction.
They conclude that the Permo-Triassic extinction consisted of two separate
extinction events: the Guadalupian event when 71 \% of marine species died
out, and the Tartarian, with an 80 \% disappearance of marine species
still the largest mass extinction in paleontological history.
The occurrence of two mass extinctions within 5 my of one another would
be possible only if the causative mechanism of the first one had ceased to
operate to allow for the observed recovery.

 The Siberian flood basalt volcanic episode occurs during the end of the
Tartarian and is a possible culprit for the Tartarian extinction.  This
volcanism commenced less than 600,000 years before the P/T boundary
much after the Guadalupian extinction. Hence the
Siberian
Traps could not have been the cause of the Guadalupian extinction. 
In addition it is likely for the 
Late Devonian extinction to also consist of two separate extinction 
episodes; the Frasnian event followed after an interval by the 
terminal Fammenian extinction [23].
The occurrence of double extinctions is explained within the 
volcanogenic dark matter framework as explained above. 
In fact this is a unique and unambiguous prediction of this model. 

Just as in the case of the earth,
dark matter capture and annihilation in other planets and their 
satellites would lead to significant 
heat generation in these bodies for a uniform dark matter halo. 
This thermal output becomes enormous when clumped dark matter 
passes through the solar system. There are several evidences
of clumpiness of dark matter in galactic halos [24].
This heat should be treated as a new source
of heat in the planetary systems, at par with primordial accretional heat
and radioactive heating. In may lie in the background or in special
circumstances manifest itself more forcefully and directly. As such 
this new source of heat in the 
solar system may lead to unique imprints. Such new signatures of 
the dark matter are found in the generation of the recent
completely unexpected discovery of the
magnetic field of Ganymede along with the enigmatic 
Mercurian magnetic field. 
Standard conventional sources of heat are unable to give a
reasonable description of these enigmatic magnetic fields.
Careful calculations within the dark matter
annihilation scenario enumerated here, explain them in 
a natural manner.

The volcanic hypothesis, despite providing a viable 
explanation for several features reported for mass 
extinctions, has always lacked a compelling 
reason for otherwise supposedly haphazard eruptions 
to occur in a periodic 
fashion. When one takes into account that the earth has been cooling 
ever since its formation ( which implies a consequent decrease 
in volcanic activity ), this objection becomes a serious
weakness. It is
hoped that a viable reason for large volcanic eruptions
to occur in a periodic manner has been presented here
with the introduction of the volcanogenic dark matter scenario. This
should strengthen the volcanic
hypothesis of mass extinctions and in addition explain the enigmatic
magnetic fields of Ganymede and Mercury. 

\newpage 
\vskip 5 mm
{\bf References}
\vskip 5 mm
1  Klapdor-Kleingrothaus H V and Staudt A,
" Non-accelerator particle physics ", IOP Publishing, Bristol
( UK ) , 1995

2  Bottino A, Donato F, Mignola G, Scopel S, Bell P and
Incichitti A, Phys Lett, {\bf B 402 } (1997) 113 

3  Glanz A, Nature, {\bf 283} (1999) 13; Cern Courier, June 1999, 17

4  Abbas S and Abbas A, Astroparticle Physics, {\bf 8} (1998) 317

5  Kanipe J, New Scientist, (Jan 11 1997) 14

6  Press W H and Spergel D N, Ap J, {\bf 296} (1985) 679

7  Krauss L M, Srednicki M and Wilczek F, Phys Rev, {\bf D33}
(1986) 2079

8  Gaisser T K, Steigman G and Tilav S, Phys Rev, {\bf D34}
(1986) 2206

9  Freese K, Phys Lett, {\bf B167} (1986) 295

10  Bengtsson H-U, Salati P and Silk J, Nucl Phys, {\bf B346}
(1990) 129

11  Silk J and Stebbins A, Ap J, {\bf 411}, 439 (1993)

12  Kolb E W and Thachev I I, Phys Rev, {\bf D50} (1994) 769

13  Bergstrom L and Ullio P, Nucl Phys, {\bf B504} (1997) 27

14  Collar J I, Phys Lett, {\bf B368} (1996) 266

15  Gould A, Ap J, {\bf 321} (1987) 571

16  Bottino A, Forengo N, Mignola G and Moscoso L,
Astroparticle Physics, {\bf 3} (1995) 65 

17  Courtillot V E, Sc Am, (Oct. 1990) 85

18  Basu A R, Poreda R J, Renne P R, Teichmann F, Vasilev 
Y R, Sobolev N V and Turrin B D, Science, {\bf 269} (1995) 822

19  Walker R J, Morgan J W, Horan M F, Science, {\bf 269} (1995) 819

20  Campbell I H, Czamanske G K, Fedorenko V A, Hill R I
and Stepanov V, Science, {\bf 258} (1992) 1760

21  Officer C B, Hallam A, Drake C L and Devine J D, Nature,
{\bf 326} (1987) 143

22  Officer C and Page J, `The Great Dinosaur Controversy',
Addison-Wesley (1996)

23  Stanley S M and Yang X, Science, {\bf 266} (1994) 1340

24  Abbas Samar, Abbas Afsar and Mohanty Shukadev,
"Evidence of Compact Dark Matter in Galactic Halos", astro-ph/9910187
 
\end{document}